\def\unit{\relax{\rm 1\kern-.29em l}}

\input harvmac
\input tables 
\input epsf.tex

\noblackbox







\lref\ISS{
  K.~Intriligator, N.~Seiberg and D.~Shih,
  ``Dynamical SUSY breaking in meta-stable vacua,''
  {\tt hep-th/0602239}.
}

\lref\Kutasov{
  D.~Kutasov,
  ``A comment on duality in ${\cal N}=1$ supersymmetric non-Abelian gauge theories,''
  Phys.\ Lett.\ B {\bf 351}, 230 (1995)
  {\tt [hep-th/9503086]}.
}

\lref\Aharony{
  O.~Aharony, J.~Sonnenschein and S.~Yankielowicz,
  ``Flows and duality symmetries in ${\cal N}=1$ supersymmetric gauge theories,''
  Nucl.\ Phys.\ B {\bf 449}, 509 (1995)
  {\tt [hep-th/9504113]}.
}

\lref\KS{
  D.~Kutasov and A.~Schwimmer,
  ``On duality in supersymmetric Yang-Mills theory,''
  Phys.\ Lett.\ B {\bf 354}, 315 (1995)
  {\tt [hep-th/9505004]}.
}

\lref\KSS{
  D.~Kutasov, A.~Schwimmer and N.~Seiberg,
  ``Chiral rings, singularity theory and electric-magnetic duality,''
  Nucl.\ Phys.\ B {\bf 459}, 455 (1996)
  {\tt [hep-th/9510222]}.
}

\lref\triangle{
  M.~J.~Duncan and L.~G.~Jensen,
  ``Exact tunneling solutions in scalar field theory,''
  Phys.\ Lett.\ B {\bf 291}, 109 (1992).
}

\lref\SQCD{
  N.~Seiberg,
  ``Electric-magnetic duality in supersymmetric nonAbelian gauge theories,''
  Nucl.\ Phys.\ B {\bf 435}, 129 (1995)
  {\tt [hep-th/9411149]}.
}

\lref\ElitzurHC{
  S.~Elitzur, A.~Giveon, D.~Kutasov, E.~Rabinovici and A.~Schwimmer,
  ``Brane dynamics and ${\cal N} = 1$ supersymmetric gauge theory,''
  Nucl.\ Phys.\ B {\bf 505}, 202 (1997)
  {\tt [hep-th/9704104]}.
}

\lref\CremadesTE{
D.~Cremades, L.~E.~Ibanez and F.~Marchesano,
``SUSY quivers, intersecting branes and the modest hierarchy problem,''
JHEP {\bf 0207}, 009 (2002)
{\tt [hep-th/0201205]}.
}



\vskip .2in  

\newbox\tmpbox\setbox\tmpbox\hbox{\abstractfont }
\Title{\vbox{\baselineskip12pt \hbox{CALT-68-2604}}}
{\vbox{\centerline{Meta-Stable Supersymmetry Breaking Vacua}  
\centerline{}
\centerline{on Intersecting Branes}}}
\vskip 0.2cm

\centerline{Hirosi Ooguri and Yutaka Ookouchi}
\bigskip
\centerline{California Institute of Technology}
\smallskip
\centerline{Pasadena, CA 91125, USA}

\vskip 1.3cm

\noindent

We identify configurations of intersecting branes
that correspond to the meta-stable supersymmetry breaking
vacua in the four-dimensional ${\cal N}=1$ supersymmetric
Yang-Mills theory coupled to massive flavors. We show
how their energies, the stability properties, and
the decay processes are described geometrically 
in terms of the brane configurations. 

\bigskip\bigskip
\Date{July, 2006}

\newsec{Introduction}

Although there is a growing body of evidences for existence of
a large landscape of string vacua, with current technology 
it is not straightforward to construct
an example of a meta-stable supersymmetry breaking vacuum
where one can demonstrate that all the moduli 
are stabilized in a controlled approximation. It is therefore 
remarkable that such vacua can be found in simple field theory 
models such as the ${\cal N}=1$
supersymmetric Yang-Mills theory coupled to massive flavors \ISS ,
suggesting that this is a generic phenomenon in field theories.
One may hope that this lesson can be applied to gravity theories 
with finite Planck mass. A possible approach toward this goal
would be to embed these field theory models in string theory
and to try to gain 
geometric insights into the existence of supersymmetry breaking
vacua.

\lref\five{
  V.~Braun, E.~I.~Buchbinder and B.~A.~Ovrut,
  ``Towards realizing dynamical SUSY breaking in heterotic model building,''
  {\tt hep-th/0606241}.
}
\lref\four{
  R.~Argurio, M.~Bertolini, C.~Closset and S.~Cremonesi,
  ``On stable non-supersymmetric vacua at the bottom of cascading theories,''
  {\tt hep-th/0606175}.
}
\lref\three{
  V.~Braun, E.~I.~Buchbinder and B.~A.~Ovrut,
  ``Dynamical SUSY breaking in heterotic M-theory,''
  {\tt hep-th/0606166}.
}
\lref\two{
  H.~Ooguri and Y.~Ookouchi,
  ``Landscape of supersymmetry breaking vacua 
in geometrically realized gauge
  theories,''
  {\tt hep-th/0606061}.
}
\lref\one{
  S.~Franco and A.~M.~Uranga,
  ``Dynamical susy breaking at meta-stable minima from 
D-Branes at obstructed  geometries,''
  JHEP {\bf 0606}, 031 (2006) 
  {\tt [hep-th/0604136]}.
}

The result of \ISS\ has recently been extended to several field 
theory models that can be constructed geometrically in string 
theory \refs{\one,\two,\three,\four,\five}. In particular,
the model studied in \two\ has a landscape of meta-stable
vacua where there are no massless scalar fields and 
the R symmetry is broken to $Z_2$. However, construction 
and analysis of the meta-stable vacua in these papers 
have been done using field theory techniques. 

In this paper, we will revisit the original model studied in
\ISS , embed it in string theory, and describe the meta-stable
vacua of the model as geometric configurations of intersecting 
branes. The energies of the meta-stable vacua are reproduced
by computing the volumes of the branes multiplied by the
brane tension, unstable modes for certain field configurations are
identified with open string tachyons, the pseudo-moduli
are stabilized by closed string exchanges, and the decay 
process of the meta-stable vacua are described as geometric
deformations of the brane configurations.

\lref\Hanany{
  A.~Hanany and E.~Witten,
   ``Type IIB superstrings, BPS monopoles, and three-dimensional gauge
   dynamics,''
   Nucl.\ Phys.\ B {\bf 492}, 152 (1997)
  {\tt [hep-th/9611230]}.
}

\lref\Giveon{
  A.~Giveon and D.~Kutasov,
  ``Brane dynamics and gauge theory,''
  Rev.\ Mod.\ Phys.\  {\bf 71}, 983 (1999)
  {\tt [hep-th/9802067]}.
}

\lref\Klemm{
  A.~Klemm, W.~Lerche, P.~Mayr, C.~Vafa and N.~P.~Warner,
  ``Self-dual strings and ${\cal N}=2$ supersymmetric field theory,''
  Nucl.\ Phys.\ B {\bf 477}, 746 (1996)
  {\tt [hep-th/9604034]}.
}

\lref\mfivezero{
  E.~Witten,
  ``Solutions of four-dimensional field theories via M-theory,''
  Nucl.\ Phys.\ B {\bf 500}, 3 (1997)
  {\tt [hep-th/9703166]}.
}

\lref\mfiveone{ 
  K.~Hori, H.~Ooguri and Y.~Oz, 
  ``Strong coupling dynamics of four-dimensional ${\cal N} = 1$
gauge theories from  M 
  theory fivebrane,''
  Adv.\ Theor.\ Math.\ Phys.\  {\bf 1}, 1 (1998) 
  {\tt [hep-th/9706082]}. 
} 
\lref\mfivetwo{
E.~Witten, 
  ``Branes and the dynamics of QCD,'' 
  Nucl.\ Phys.\ B {\bf 507}, 658 (1997)
  {\tt [hep-th/9706109]}. 
} 

\lref\mfivethree{
  A.~Brandhuber, N.~Itzhaki, V.~Kaplunovsky, J.~Sonnenschein and S.~Yankielowicz,
  ``Comments on the M theory approach to ${\cal N} = 1$ SQCD and brane dynamics,''
  Phys.\ Lett.\ B {\bf 410}, 27 (1997)
  {\tt [hep-th/9706127]}.
}

The construction of the brane configuration in this
paper follows the approach initiated in \Hanany\
and applied to the model relevant to this paper in 
\ref\ElitzurFH{
  S.~Elitzur, A.~Giveon and D.~Kutasov,
  ``Branes and ${\cal N} = 1$ duality in string theory,''
  Phys.\ Lett.\ B {\bf 400}, 269 (1997)
  {\tt [hep-th/9702014]}.
}. For an extensive review of
this approach, see \Giveon . The intersecting brane construction
is related by a chain of duality to the geometric engineering
initiated in \Klemm . It is likely that most of what
we will discuss in this paper can be stated in the language
of local Calabi-Yau geometry by reverse engineering the
latter approach. The geometric engineering description 
of the meta-stable vacua may involve non-K\"ahler geometries,
and study in this direction may lead to new insights
into string theory on such non-supersymmetric geometries. 
It may also be possible to uplift
our result to M theory, where D4 and NS5 branes become 
M5 branes and D6 branes are replaced by the Taub-NUT geometry
\refs{\mfivezero, \mfiveone, \mfivetwo, \mfivethree},
along the line of the approach in \ref\deBoerBY{
  J.~de Boer, K.~Hori, H.~Ooguri and Y.~Oz,
  ``Branes and dynamical supersymmetry breaking,''
  Nucl.\ Phys.\ B {\bf 522}, 20 (1998)
  {\tt [hep-th/9801060]}.
}.

\newsec{Brane configurations for the meta-stable vacua} 

We propose a configuration of intersecting branes
in type IIA string theory which correspond to 
the meta-stable vacuum with broken supersymmetry
in the ${\cal N}=1$ $U(N_c)$ super Yang-Mills theory coupled to 
$N_f$ chiral multiplets, which we will refer to as quarks,
in the fundamental representation of $U(N_c)$. 
Throughout of this paper, we assume $N_f > N_c$.

\bigskip

\leftskip 2pc \rightskip 2pc
\bigskip
\centerline{\epsfxsize 5.2truein\epsfbox{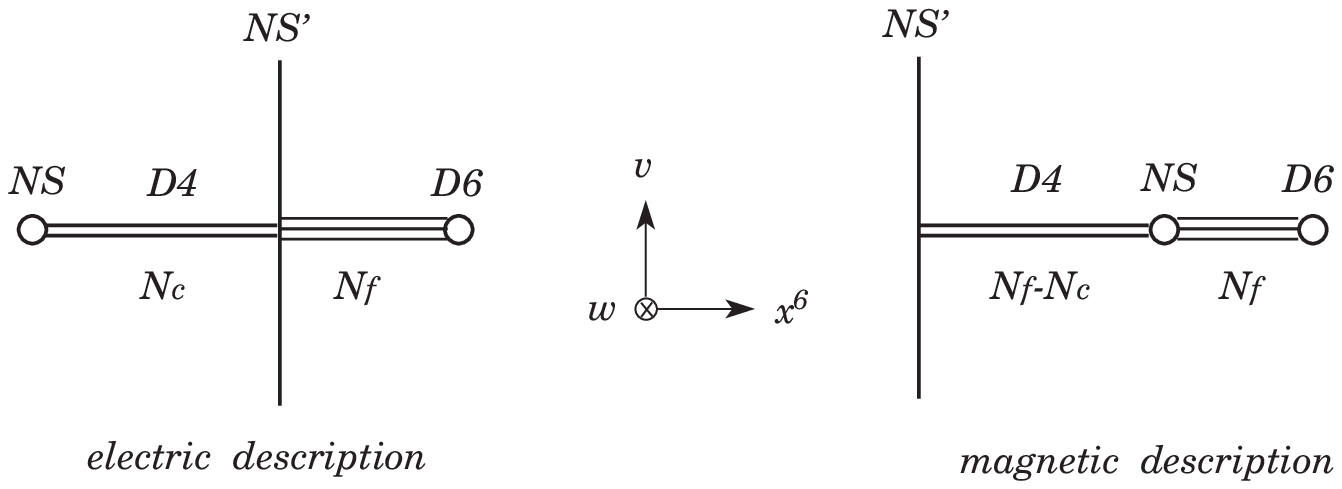}}
 \noindent{\ninepoint\sl \baselineskip=2pt {\bf
Fig.1} {{The supersymmetric brane configurations for 
the ${\cal N}=1$ supersymmetric gauge theory with massless
flavors. The vertical axis represents the holomorphic
coordinate $v = x^4 + ix^5$, and the horizontal axis is for
$x^6$. The $w = x^7 + ix^8$ and $x^9$ coordinates
are suppressed in these diagrams. All the branes share
the (0123) plane, where the four-dimensional
gauge theory is defined.}}}
\bigskip
\leftskip 0pc \rightskip 0pc
 
\bigskip

In the electric description, the theory can be realized on
the network of branes consisting of the following:

\smallskip

\noindent
 \item{$\bullet$} One NS5 brane stretched in the (0123) and (78) directions and
located at $v, x^6, x^9=0$, where $v=x^4+ix^5$. 
We call this as the NS brane. 

\item{$\bullet$} One NS5 brane stretched in the (0123) and (45) directions and
located at $w, x^9=0$ and $x^6=L$, where
$w=x^7+ix^8$ and $L>0$. We call this
as the NS' brane. 
 
\item{$\bullet$} $N_f$ D6 branes stretched 
in the (0123) and (789) directions and
located at $v= m_i$ and
$x^6 = L' > L$. Here $m_i$ with $i=1,..., N_f$
are complex numbers that are identified
with the masses of the quarks. 

\item{$\bullet$} $N_c$ D4 branes stretched in the (0123) directions and 
going between the NS and NS' branes along the $x^6$ axis. They
are located at $v, w, x^9=0$.

\item{$\bullet$} $N_f$ D4 branes extended in the (0123) directions
and going between the NS' brane at $x^6=L$ to 
the D4 branes at $x^6=L'$ along the $x^6$ axis. They
are located at $v=m_i$ and $w, x^9=0$. 

\lref\bachastwo{
  C.~Bachas and M.~B.~Green,
  ``A classical manifestation of the Pauli exclusion principle,''
  JHEP {\bf 9801}, 015 (1998)
  {\tt [hep-th/9712187]}.
}

\lref\bachasone{
  C.~P.~Bachas, M.~B.~Green and A.~Schwimmer,
  ``(8,0) quantum mechanics and symmetry enhancement in type I'
  superstrings,''
  JHEP {\bf 9801}, 006 (1998)
  {\tt [hep-th/9712086]}.
}

\lref\ov{
  H.~Ooguri and C.~Vafa,
  ``Geometry of ${\cal N} = 1$ dualities in four dimensions,''
  Nucl.\ Phys.\ B {\bf 500}, 62 (1997)
  {\tt [hep-th/9702180]}.
}

\smallskip
\noindent
The $s$-rule of \Hanany\ states that
it is not possible to suspend more than one D4 branes between 
one NS5 brane and one D6 brane while maintaining 
supersymmetry.\foot{In \Hanany, the $s$-rule was postulated 
to reproduce moduli spaces of gauge theories
correctly. Subsequently this rule has been derived 
from various points of view in \refs{\ov,  \mfiveone,
\bachasone, \bachastwo}.}
Thus, we need a D6 brane for each one of $N_f$ D4 branes.
The D4 branes are located at $v=m_i$,  and
they are parallel to the $N_c$ D4 branes between the NS
and NS' branes as required by supersymmetry. The
resulting brane configuration,  
when all the quarks are massless, $m_i=0$, 
is shown on the left-side of Figure 1.

The massless sector of open strings going between the $N_c$ 
D4 branes gives rise to the vector multiplet for the gauge 
group $U(N_c)$.
In addition, $N_f$ flavors of quarks $(Q_i, \tilde{Q}_i)$ 
with masses $m_i$  arise from open strings between 
the $N_c$ D4 branes and the $N_f$ D4 branes. The open strings
on the $D6$ branes decouple since the $D6$ branes have infinite
volumes in the (789) directions. Thus, the low energy
effective theory on the
branes is the supersymmetric Yang-Mills theory
couples to $N_f$ flavors on the 4-dimensional plane
in the (0123) directions shared by all the branes. It was argued in 
\mfivezero\ that an infrared singularity on the NS5 branes freezes 
the diagonal $U(1)$ factor in the $U(N_c)$ group.
In the following, we will discuss as if
the gauge group is $U(N_c)$ since the $U(1)$ factor 
is infrared free and decouples from 
the $SU(N_c)$ dynamics  in any case.  

The brane configuration has $U(1)_{78}$ 
global symmetry corresponding to the phase rotation of
the coordinate $w=x^7+ix^8$. This is identified with
the R symmetry in the gauge theory. 
In addition, if all quark masses are zero,
we have $U(1)_{45}$ symmetry corresponding to the phase 
rotation in the coordinate $v=x^4+ix^5$. In this case, there is 
also the flavor $U(N_f) \times U(N_f)$ symmetry, where the diagonal
$U(N_f)$ is generated by exchanges of the D6 branes.

\lref\Seibergone{
  N.~Seiberg,
  ``Exact results on the space of vacua of four-dimensional SUSY gauge
  theories,'' 
  Phys.\ Rev.\ D {\bf 49}, 6857 (1994);
  {\tt hep-th/9402044}. 
}
 
\lref\Seibergtwo{
  N.~Seiberg,
  ``Electric-magnetic duality in supersymmetric nonAbelian gauge theories,''
  Nucl.\ Phys.\ B {\bf 435}, 129 (1995);
  {\tt hep-th/9411149}.
}

The brane configuration for the magnetic dual for this
theory was identified in \ElitzurFH . Let us assume for the 
moment that all the quark masses are zero. 
To go from the electric description to the magnetic dual,
one exchanges the locations of the NS and NS' branes
in the (69) plane. When $N_f > N_c$, the resulting configuration 
consists of:

\smallskip

\item{$\bullet$} NS' brane at $w, x^6, x^9=0$.

\item{$\bullet$} NS brane at $v, x^9=0$ and $x^6 = L''$ with
$0 <  L'' < L'$.

\item{$\bullet$} $N_f$ D6 branes at $v=0$ and $x^6=L'$.
(We are setting $m_i=0$.)

\item{$\bullet$} $(N_f - N_c)$ D4 branes between
the NS' and NS branes.

\item{$\bullet$} $N_f$ D4 branes between the NS brane and D6 branes. 

\smallskip
\noindent
The low energy physics is 
described by the $U(N_f-N_c)$ vector multiplet
coupled to $N_f$ quarks $(q_i, \tilde q_i)$ in 
the fundamental representation,
and a gauge neutral meson $M_{ij}$,
which transforms in the adjoint representation in the
$U(N_f)$ flavor symmetry. The extra meson degrees of
freedom correspond to the motion of the $N_f$ D4 branes
in the $w=x^7+ix^8$ directions along the NS and D6 branes. 
The superpotential $W$ in the magnetic description is
proportional to $\tr \ \tilde{q} M q$,
where $\tr$ is over the fundamental representation of 
$U(N_f-N_c)$ and the sum over the flavor indices is
implicit. This reproduces the magnetic dual of the
theory identified by Seiberg \refs{\Seibergone, \Seibergtwo}.

The meson field $M_{ij}$ is identified with the bilinear
combination $Q_i \tilde{Q}_j$ of the quarks in the 
electric description. Thus, turning on the quark masses
in the electric description corresponds to deforming
the superpotential by adding a term linear in $M$ as
\eqn\superpot{ W = \tr\ \tilde{q} M q +
 \tr'\ m M,}
where $\tr$ and $\tr'$ are the traces over the gauge
and the flavor indices respectively and $m$ is the
$N_f \times N_f$ quark mass matrix.\foot{We assume 
that $m$ obeys $[m, m^\dagger]=0$ so that it is diagonalizable 
with the eigenvalues given by the quark masses $m_i$.} 
The $F$-term conditions are
\eqn\fterm{ \eqalign{ &q \tilde q + m = 0,\cr
   & M q = 0, ~~ \tilde{q} M = 0.}}
Since the ranks of $q$ and $\tilde q$ are at most
$(N_f-N_c)$, the first equation cannot be satisfied
if the rank of the $N_f \times N_f$ mass matrix $m$ 
exceeds $(N_f-N_c)$. In particular, turning
on masses for all the quarks breaks the supersymmetry
in the magnetic description. This is the rank
condition mechanism of \ISS.

\subsec{Brane configurations for the meta-stable vacua}

For simplicity, let us use the flavor symmetry
to arrange the mass matrix $m$ so that
\eqn\massmatrix{ m = {\rm diag}(m_1, m_2, \cdots ,
m_{N_f}),}
with
\eqn\order{ |m_1| \geq |m_2| \geq \cdots \geq |m_{N_f}|.}
The local minima of the tree-level $F$ and $D$-term 
potential can be parametrized as
\eqn\localmin{\eqalign{&
q = \pmatrix{ \varphi_0 \cr 0}, ~~
\tilde q = \pmatrix{\tilde \varphi_0, & 0 } \cr
& M = \pmatrix{ 0 & 0 \cr 0 & M_0 },}}
where $\varphi_0$ and $\tilde\varphi_0$ 
are $(N_f - N_c) \times (N_f - N_c)$ matrices
satisfying
\eqn\whatvarphi{ \varphi_0 \tilde\varphi_0 = -
{\rm diag}(m_1, m_2, \cdots, m_{N_f-N_c}), }
and $M_0$ is an arbitrary $N_c \times N_c$ matrix. 
It is important that we choose the $(N_f-N_c)$ largest
masses for the eigenvalues of $q\tilde q$.
Otherwise, the configuration is unstable under small perturbations
\ISS. We will show how this instability 
effect can be seen in the brane configuration.

In the brane configuration, turning 
on the quark masses corresponds to moving
the D6 branes from $v=0$ to $v=m_1,..., m_{N_f}$.
Suppose we turn on masses for $n \leq (N_f-N_c)$ quarks.
To move the $n$ D6 branes while maintaining supersymmetry, 
we need to connect $n$ pairs of 
D4 branes across the NS brane at $x^6=L''$ and move them together
with the D6 branes since the D4 branes must be parallel to
each other. The resulting configuration contains:

\smallskip

\item{$\bullet$} $n$ D4 branes stretched between the NS' and D6 branes
at $v = m_i$ with $i=1,...,n$.

\item{$\bullet$} $(N_f-N_c-n)$ D4 branes between the NS' and NS branes.

\item{$\bullet$} $(N_f-n)$ D4 branes between the NS and D6 branes
at $v=0$.

\smallskip
\noindent
Since the D6 branes impose the Dirichlet boundary condition
on the (0123) components of the gauge fields on the 
D4 branes, the gauge symmetry is broken
to $U(N_f-N_c-n)$. This is consistent with the field
theory fact that the quark bilinear $q\tilde q$ gets
a vacuum expectation value of rank $n$ and spontaneously
break the gauge symmetry.  

This works until $n$ hits $(N_f-N_c)$ when one finds that 
there is no more D4 brane left between the NS' and NS
branes. If we add mass terms for more quarks
and move the corresponding D6 branes away from 
the origin of the $v$ plane, the D4 branes connecting them 
with the NS brane will have to be tilted
in the $v$-$x^6$ plane. In particular, these D4 branes 
will not be parallel with the $(N_f-N_c)$ D4 branes 
going along the $x^6$ axis. See Figure 2. 

\bigskip

\leftskip 2pc \rightskip 2pc
\bigskip
\centerline{\epsfxsize 3.5truein\epsfbox{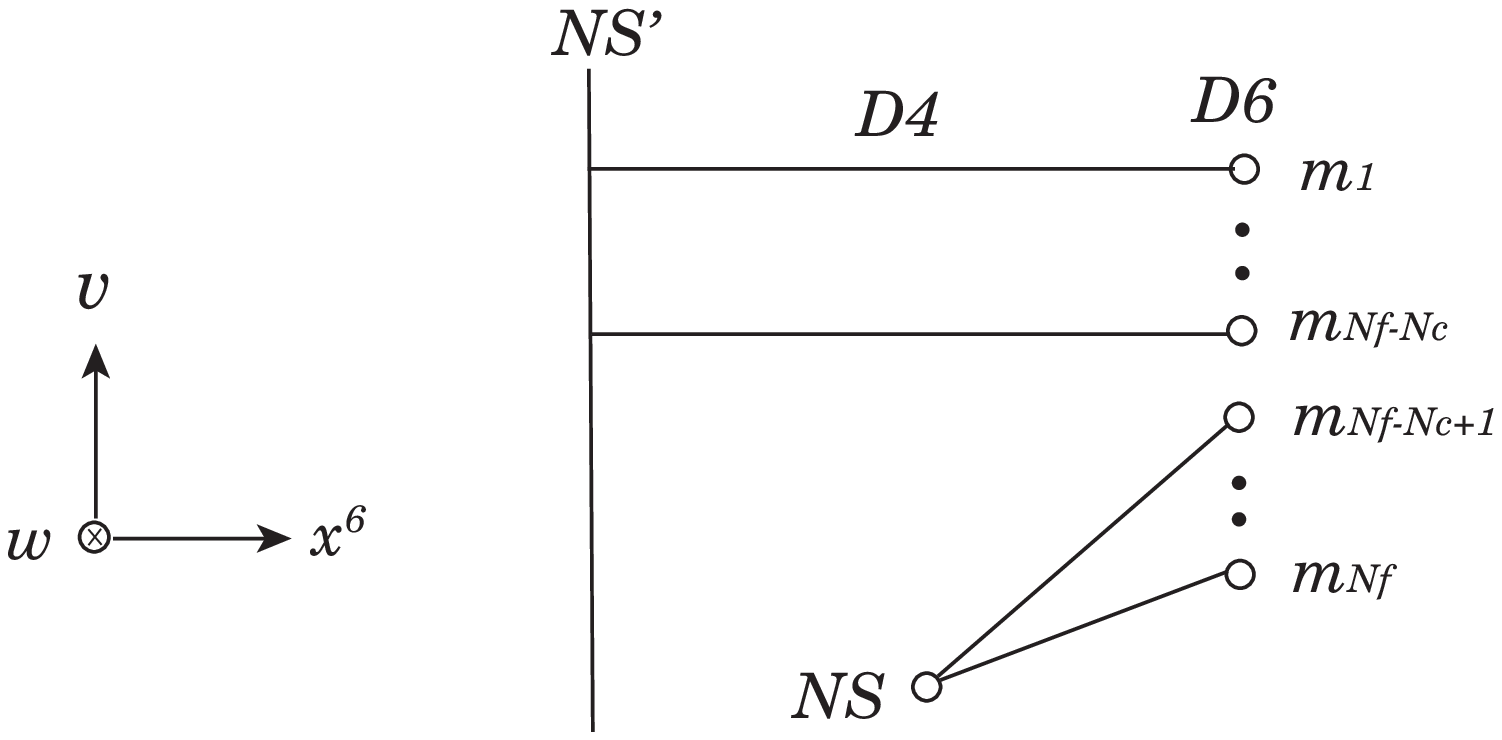}}
  \noindent{\ninepoint\sl \baselineskip=2pt {\bf
Fig.2} {{The brane configuration for the meta-stable 
supersymmetry breaking vacuum. The horizontal and
vertical directions represent the $x^6$ and $v$ coordinates
respectively. 
The $w$ and $x^9$ directions are suppressed in this diagram.}}} 
\bigskip
\leftskip 0pc \rightskip 0pc

\bigskip

We claim that 
the resulting brane configuration corresponds to the field
configuration given by  \localmin\ and \whatvarphi .
Note that the quark 
bilinear $-q \tilde q$ specifies the $v$ coordinates of 
the D4 branes at $x^6=L''$. This follows from the fact that
 the quarks in the magnetic description
are defined as open strings going between the two types of D4
branes across the NS brane, which is located at $x^6=L''$. 
From the brane configuration, we can read off that 
$$-q \tilde q = {\rm diag}(m_1,...,m_{N_f-N_c},0,...,0)$$
since the $(N_f-N_c)$ D4 branes
have been moved to $v=m_1,...,m_{N_f-N_c}$ along with the D6 branes
while the $N_c$ D4 branes are still ending on the NS brane
at $v=0$.
Since these $(N_f-N_c)$ D4 branes are frozen at $w=0$ while 
the $N_c$ D4 branes can be moved to any locations 
along the $w$ direction at the string tree level, 
the expectation value of the meson $M$ can be expressed as
$$M = \pmatrix{ 0 & 0 \cr 0 & M_0 }, $$
with an arbitrary $N_c \times N_c$ matrix $M_0$.
This reproduces the field configuration given by \localmin\
and \whatvarphi . The R symmetry
of the gauge theory is realized as the phase rotation of the
$w$ coordinate. It is clear that the brane configuration  
corresponding to the meta-stable vacuum preserves this R symmetry
when $M_0=0$ since all the branes are invariant under the rotation. 
In the following, we will show that this brane configuration
captures various other features of the meta-stable vacuum. 

\subsec{Tachyons}

Since the brane configuration proposed in the above
does not preserve supersymmetry, we need to make sure that 
the configuration is locally stable. Let us verify that 
there are no tachyons in the open string spectrum. 

Extrema of the tree-level potential 
is parametrized as \localmin\ with $\varphi_0$
given by \whatvarphi . Suppose we did not order the
masses as in \order, and $|m_{N_f}| > |m_1|$ for example.
One can show that, in this case, fluctuations of the 
dual quarks $(q,\tilde q)$ at $M=0$ contain tachyonic modes with
(mass)$^2$ given by
\eqn\tachyonmass{ {\rm (mass)}^2 = -
{  |m_{N_f}| - |m_1| \over Z \widehat{\Lambda}},}
where $Z$ is the normalization factor in 
front of the kinetic 
term of the meson field $M$ in the magnetic theory
when the superpotential is normalized as  
$\tr \ m M + \cdots$, and $\widehat{\Lambda}$ is 
some mass parameter of the magnetic dual of the gauge
theory that cannot be determined by the information in
the electric theory alone (see the comment below (5.6)
in \ISS). Thus, the tree level stability of the field
configuration at $M=0$ requires that the quark masses 
should be lined up as in \order . 

\bigskip

\leftskip 2pc \rightskip 2pc
\bigskip
\centerline{\epsfxsize 4.5truein\epsfbox{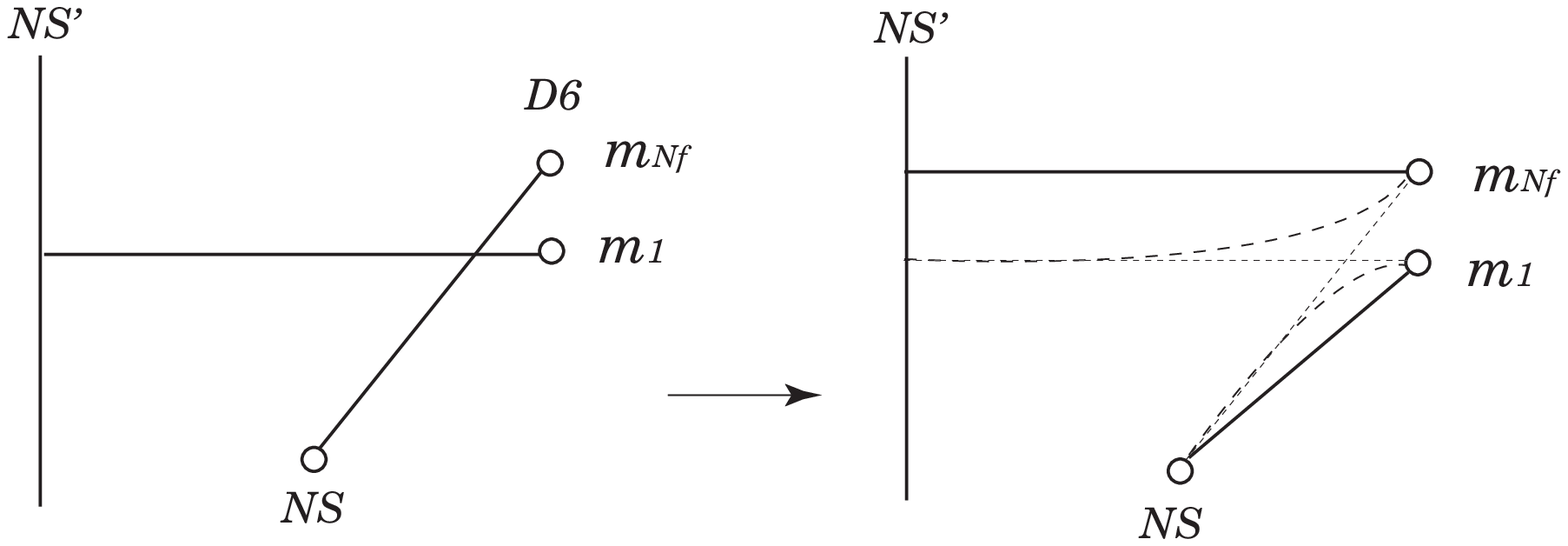}} 
\noindent{\ninepoint\sl \baselineskip=2pt {\bf
Fig.3} {{When $m_{N_f} > m_1$, the two D4 branes intersect.
The tachyon condensation recombines the two branes.}}} 

\leftskip 0pc \rightskip 0pc

\lref\intersectone{
  M.~Berkooz, M.~R.~Douglas and R.~G.~Leigh,
  ``Branes intersecting at angles,''
  Nucl.\ Phys.\ B {\bf 480}, 265 (1996)
  {\tt [hep-th/9606139]}.
}
\lref\intersecttwo{
  A.~Hashimoto and W.~Taylor,
  ``Fluctuation spectra of tilted and intersecting D-branes from the
  Born-Infeld action,''
  Nucl.\ Phys.\ B {\bf 503}, 193 (1997)
  {\tt [hep-th/9703217]}.
}
\lref\intersectthree{
  K.~Hashimoto and S.~Nagaoka,
   ``Recombination of intersecting D-branes 
  by local tachyon condensation,''
  JHEP {\bf 0306}, 034 (2003)
  {\tt [hep-th/0303204]}.
}
\lref\intersectfour{
  F.~T.~J.~Epple and D.~Lust,
  ``Tachyon condensation for intersecting branes at small and large angles,''
  Fortsch.\ Phys.\  {\bf 52}, 367 (2004)
  {\tt [hep-th/0311182]}.
}

\bigskip

This instability can be seen in the brane configuration
as follows. For simplicity, let us assume that 
all the masses are real. The first $(N_f-N_c)$ masses 
$m_1, ... , m_{N_f-N_c}$ are the location of the D6 branes
which are connected to the NS' brane through D4 branes. 
The remaining $N_c$ masses $m_{N_f-N_c+1},...,m_{N_f}$
are the locations of the D6 branes connected to
the NS brane through D4 branes. 
The latter D4 branes are in angles in the $v$-$x^6$ plane and
the angles are determined by the quark masses.
Suppose $m_1 < m_{N_f}$ for example. 
As we can see in the left-side of Figure 3, 
the D4 branes connected to the
D6 branes at $m_1$ and $m_{N_f}$ intersect when both branes
are located at $w=0$. 
It is well-known that, when a pair of branes are intersecting
at an angle as described in the above, the spectrum of the
open string stretched between the branes contains a tachyon
and that the end point of the tachyon condensation is
a recombination of the branes \refs{\intersectone,
\intersecttwo, \intersectthree, \intersectfour} as in the
right-side of Figure 3. After the
recombination, the D6 brane at $m_{N_f}$ is connected to
the NS' brane through a D4 brane 
and the D6 brane at $m_1$ is connected to the NS brane
brane through a D4 brane at an angle in the $v$-$x^6$ plane. 
After a series of 
recombinations, the brane configuration will settle down 
to the configuration that corresponds to \localmin\ in 
the field theory with the quark masses ordered as 
$m_1 \geq \cdots \geq  m_{N_f}$. This matches well with
what we expect in the field theory.

If the tachyonic mode in the field theory is to be identified
with the open string tachyon at the intersection of the
two D4 branes, we should be able to
understand the mass formula \tachyonmass\ from the string theory
point of view also. The mass of the open string tachyon 
on the branes at an angle $\theta$ is given by \intersectone 
\eqn\tachyonangle{ {\rm (mass)}^2 = -{|\theta| \over l_s^2}. }
To compute the angle $\theta$, we note that 
one of the D4 branes is at $v=m_1$
and the other is going from $v=0$ to $v=m_{N_f}$. 
The physical distance for the separation of the two
end points of the second D4 brane when projected on the
$v$ plane is $m_{N_f} l_s^2$. On the other hand
the length of the D4 brane is proportional to the
normalization factor $Z$ of the meson kinetic term.
Since the meson field $M$ is identified with the quark 
bilinear $Q \tilde Q$ in the electric variable, 
the factor $Z$ should have the dimensions of (length)$^2$.
Thus, the angle $\theta$ of the D4 brane can be expressed as
\eqn\angleapprox{
 \tan \theta = {m_{N_f} l_s^2 \over Z \Lambda'}  ,}
with some dimensionful parameter $\Lambda'$.
With this, the tachyon mass in the field theory limit
$l_s \rightarrow 0$ is estimated as
\eqn\tachyonformula{ {\rm (mass)}^2 = - {|\theta| \over l_s^2}
\rightarrow -  {m_{N_f}\over Z \Lambda'} . }
This reproduces the first half of the
tachyon mass formula \tachyonmass\ if we identify
$\Lambda'$ with the unknown parameter $\widehat{\Lambda}$ 
in the magnetic theory. 

The above formula \tachyonformula\ is obtained by ignoring
the fact that the D4 branes end on the D6 branes.
We claim that the second half of the tachyon mass
formula \tachyonmass\ is due to
the D6 brane boundary condition. The formula shows
in particular that 
the tachyon becomes massless in the limit of
$m_{N_f}=m_1$.
Let us try to understand this phenomenon by taking into
account the boundary conditions at the D6 branes. 

According to \intersectthree , if we take the field theory limit
$l_s \rightarrow 0$ so that $\theta/l_s^2$ remains finite 
as in our case, 
the string theory tachyon can be described as a classical 
configuration of the $U(2)$ gauge theory along the $x^6$ 
direction as follows. Consider the gauge field $A_6$ 
in the $x^6$ direction and the scalar field $\phi$ in 
the adjoint of $U(2)$ corresponds to the oscillation
of the branes in the $v$ direction. The fact that the
branes intersect at the angle $\theta$ 
can be expressed
in the gauge theory language as the expectation value
of the scalar field $\Phi$ as follows, 
\eqn\gaugeangle{ \Phi(x^6) = \sigma^3   {\theta \over l_s^2}
x^6,}
where we assume that the intersection is at $x^6=0$
and $\sigma^{1,2,3}$ are the Pauli matrices. 
We can evaluate the fluctuation spectrum of the D4 branes
around this background
by expanding the gauge theory action in quadratic order
in $A_6$ and $\Phi$. 
If we ignore the boundary conditions at the NS and $D6$ branes,
the gauge theory equations of motion combined with the
Gauss law constraint for the gauge $A_0=0$ imply that
the lowest energy excitation is of the form,
\eqn\tachyongauge{ \eqalign{& A_6 =   \sigma^2  \exp \left[
- {|\theta| \over l_s^2} (x^6)^2\right], \cr
& \delta \Phi =  \sigma^1  \exp\left[
- {|\theta| \over l_s^2} (x^6)^2\right].}}
There is another excitation with the same energy
obtained by exchanging $(\sigma^1, \sigma^2) \rightarrow
(-\sigma^2, \sigma^1)$. The (mass)$^2$ for this
configuration correctly reproduces the formula \tachyonangle .

In the $A_0=0$ gauge, the D6 branes impose the
Neumann boundary condition on $A_6$ and the Dirichlet
boundary condition on $\delta \Phi$. Since 
\tachyongauge\ does not satisfy these conditions, we expect
that the tachyon (mass)$^2$ is raised by the
boundary condition. The distance of the intersection
point $x^6=0$ to the boundary at the D6 branes 
is proportional to 
$(m_{N_f} - m_1)$, and the effect of the boundary condition
should become greater as $m_{N_f}$ approaches $m_1$. 
In the limit of $m_{N_f} = m_1$, the tachyon becomes
massless. This can be verified directly by solving
the gauge theory equations combined with the Gauss law
constraint.
When $m_{N_f} = m_1$, the boundary condition
has the global $U(2)$ symmetry for the exchange 
of the two D6 branes, but it is spontaneously broken
by the field expectation value \gaugeangle . This gives rise
to the Nambu-Goldstone boson, and that is the massless
mode that appears in the limit of the tachyon at $m_{N_f}
= m_1$. 
Note that this Nambu-Goldstone mode would be non-normalizable
if the branes were infinitely extended since $\delta \Phi$ grows
linearly in $x^6$, but it is normalizable on
our D4 branes of finite lengths and we should count
it in the spectrum. We have also examined other
normalizable modes on the half line $x^6 \leq 0$.\foot{We 
thank K. Hashimoto on communication on
this point and for sharing his unpublished notes with us,
which were useful for our analysis here.}
The Dirichlet boundary condition
on $\delta \Phi$ at $x^6=0$ can be taken into account by extending 
the space to $x^6 > 0$ and by requiring that $\delta \Phi$ 
be odd under the reflection of $x^6 \rightarrow -x^6$. 
We found that all the fluctuations 
other than the Nambu-Goldstone modes have positive
(mass)$^2$.
Thus, in the limit of $m_{N_f} = m_1$,
the open string tachyon is removed by the D6 brane
boundary condition and the brane configuration becomes
stable. 

If we move the D6 branes further so that $m_1 > m_{N_f}$, 
the Nambu-Goldstone mode becomes massive as the open
string between the D4 branes are separated even at
the end points of the D4 branes at the D6 branes. 
It is straightforward to check that there are no other
sources of open string tachyons in this configuration
when $m_1 \geq \cdots \geq m_{N_f}$ even though the 
brane configuration breaks 
supersymmetry. To our knowledge, this way 
of eliminating tachyons has not been considered
in phenomenological model building based on intersecting
branes.\foot{This is different from the quasi-supersymmetric
construction studied in \CremadesTE , where each intersection
preserves some supersymmetry but 
the whole configuration breaks supersymmetry.
In the present case, when the quark masses are degenerate,
there are D4 branes meeting with angle, breaking all
supersymmetry. The potential tachyon is eliminated by
the boundary condition at the D6 branes.}
It would be interesting to explore possibilities of 
superstring model building using configurations like this. 

\subsec{Vacuum energy} 

The energy of the meta-stable vacuum is higher than
that of the supersymmetric brane configuration at $m_i=0$ 
since the D4 branes at angles are longer. The angles are 
$$ \theta_i  \sim  {|m_i| \over Z}  , 
~~~~i=N_f-N_c+1, ..., N_f, $$
and the length of the D4 brane times 
the brane tension is proportional to 
the normalization factor $Z$ of the meson kinetic term.
Thus, if we set the energy of the supersymmetric
brane configuration to be zero, the energy density $V$ for
the meta-stable configuration can be estimated as
\eqn\vacenergy{ V= \sum_{i=N_f-N_c+1}^{N_f} 
\left({1 \over \cos \theta_i} - 1\right)  Z
 \ \sim \sum_{i=N_f-N_c+1}^{N_f} \theta_i^2 Z
 \ \sim \ {1 \over Z} \sum_{i=N_f-N_c+1}^{N_f} |m_i|^2.}
This agrees with the value of the 
$F$-term potential 
$$ V = {1\over Z} \left| {\partial W \over \partial M}\right|^2, $$
evaluated for the field configuration \localmin .

\subsec{Pseudo-moduli and their one-loop effective potential} 

At the tree level in the field theory analysis, there is
no potential for deformation of $M_0$ in \localmin . 
Thus, $M_0$ is called pseudo-moduli. Since there
are non-compact directions in $M_0$, it is important
to find out if these directions are stabilized by
quantum effects. 

The pseudo-moduli $M_0$ describe locations
of the $N_c$ D4 branes in the $w$ direction along the 
NS and D6 branes. In the field theory, 
it was shown in \ISS\ that 
the one-loop Coleman-Weinberg potential lifts these
flat directions, except for those protected by the Goldstone
theorem. In the string picture, the one-loop computation
in the gauge theory is the $l_s \rightarrow 0$ limit
of the one-loop open string computation. By the worldsheet
duality, it is related to exchange of closed strings. If
the branes are not in angles, the closed string exchange
does not generate a potential
because of the cancellation of effects due to exchanges of NS-NS
states and RR states. Since the $(N_f-N_c)$ D4 branes and
the $N_c$ D4 branes are at angles in our case,
the cancellation is not perfect. Thus we expect that a potential
is generated for $M_0$. The one-loop field theory analysis
predicts that this potential is attractive for all the
non-compact directions in $M_0$. 

When the $N_c$ D4 brane segments are far from the other
$(N_f-N_c)$ D4 branes, main contributions to the potential
come from graviton and RR fields exchange. Since the
branes are at angles, the effect of the RR field exchange,
which is repulsive, is weaker than the effect of the graviton
exchange, which is attractive. Thus, the potential should
be attractive when the two sets of D4 branes are widely 
separated. This is consistent with the field theory analysis.
Unfortunately,
this is not in the field theory regime where
the distances between the D4 branes are less than
$l_s$. It would be useful to develop general criteria
in the language of brane configurations 
to decide when a potential between branes at angles
becomes attractive so that we can tell when supersymmetry
breaking configurations are locally stable. 

\subsec{Decay of the meta-stable vacua}

In addition to the meta-stable supersymmetry breaking vacuum, 
the theory has supersymmetric vacua. We will show
how the decay of the meta-stable vacuum into 
the supersymmetric vacua is described in
the brane construction.

\bigskip

\leftskip 2pc \rightskip 2pc
\bigskip
\centerline{\epsfxsize 4.5truein\epsfbox{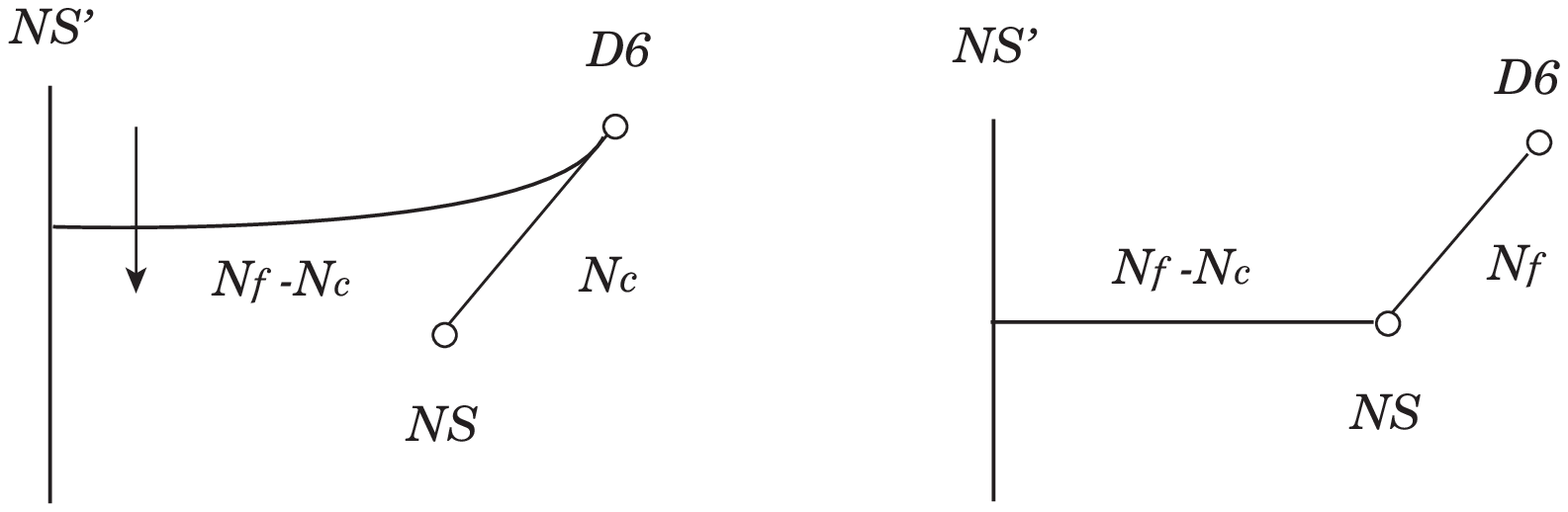}} 
\noindent{\ninepoint\sl \baselineskip=2pt {\bf
Fig.4} {{In order for the meta-stable vacuum to decay into 
supersymmetric vacua, it has to climb up the potential
barrier. In the brane construction, it is described as
the bending of the $(N_f-N_c)$ D4 branes.}}} 
\bigskip
\leftskip 0pc \rightskip 0pc

To go from the meta-stable vacuum to the supersymmetric
vacua, we first move the $(N_f-N_c)$ D4 branes down toward
the NS brane so that we end up having $(N_f-N_c)$ D4 branes
connecting the NS' and NS branes and $N_f$ D4 branes connecting
the NS and D6 branes. See Figure 4.
This process costs energy as the 
$(N_f-N_c)$ D4 branes are bent and their lengths increase.  
We can estimate the extra energy density $\Delta V$ for this
configuration as
\eqn\extracost{ \Delta V \sim \sum_{i=1}^{N_f-N_c} 
\left({1 \over \cos \theta_i} - 1\right)  Z
\ \sim \ {1\over Z} \sum_{i=1}^{N_f-N_c} |m_i|^2. }
In the field theory, this corresponds to 
$q = \tilde q = 0$ and $M=0$.
This configuration was considered in 
\ISS\ as an intermediate state in the decay of the 
meta-stable vacuum. The $F$-term potential evaluated
for this field theory configuration agrees 
with \extracost .

\lref\massmfiveone{
  M.~Schmaltz and R.~Sundrum,
  ``${\cal N} = 1$ field theory duality from M-theory,''
  Phys.\ Rev.\ D {\bf 57}, 6455 (1998)
  {\tt [hep-th/9708015]}.
}

\lref\massmfivetwo{
  S.~Sugimoto,
  ``Comments on duality in MQCD,''
  Prog.\ Theor.\ Phys.\  {\bf 100}, 123 (1998)
  {\tt [hep-th/9804114]}.
}

\lref\massmfivethree{
  K.~Hori,
  ``Branes and electric-magnetic duality in supersymmetric QCD,''
  Nucl.\ Phys.\ B {\bf 540}, 187 (1999)
  {\tt [hep-th/9805142]}.
}

\lref\massmfivefour{
  J.~Evslin, H.~Murayama, U.~Varadarajan and J.~E.~Wang,
  ``Dial M for flavor symmetry breaking,''
  JHEP {\bf 0111}, 030 (2001)
  {\tt [hep-th/0107072]}.
}

Since there are $N_f$ D4 branes connecting the NS
and D6 branes, we can move them in the $w$ direction.
The locations of the D4 branes in the $w$ plane are
specified by the expectation value of $M$. 
The field theory result shows that 
there are supersymmetric brane configurations at 
\eqn\susym{ M_{ij} = m^{-1}_{ij} (\det m)^{{1\over N_c}}
    \left( \Lambda^{3N_c-N_f}\right)^{{1\over N_c}}, }
where $\Lambda$ is the strong coupling scale of the
electric theory. For a generic mass matrix $m$ with 
rank $m = N_f$, the $N_f$ D4 branes are all separated
and away from the origin of the $w$ plane. This is why
the brane configuration needs to climb up the potential
barrier, reach the stage shown in Figure 4, and
let all the $N_f$ D4 brane segments be moved away,
before relaxing itself to the supersymmetric configurations.

We can also derive the locations of the $N_f$ D4 branes 
in the supersymmetric vacua by lifting the brane configurations 
to M theory. The description of the supersymmetric vacua in
the M theory has already been given 
in \refs{\mfivezero , \mfiveone , 
\mfivetwo, \mfivethree} and we will not repeat the
analysis here. In the M theory description, 
the NS5 brane and D4 branes are interpreted as M5 branes
and the D6 branes are replaced by the Taub-NUT geometry. 
The M5 brane configurations are supersymmetric if they
are holomorphic. From the subsequent analysis of the M5
brane configurations in 
\refs{\massmfiveone, \massmfivetwo, \massmfivethree, 
\massmfivefour}, it is clear that they can be 
interpreted as the M theory lift of the D4/D6/NS5 brane
configurations that can be reached by moving the $N_f$ 
D4 branes along the NS and D6 branes starting from the 
configuration in Figure 4. This is how the meta-stable
brane configuration decays into the supersymmetric configurations.

\bigskip
\bigskip

While the manuscript of this paper is being finalized, 
we were informed by Ofer Aharony of a related work in 
progress by Shimon Yankielowicz.

\bigskip

\centerline{\bf Acknowledgments}
\bigskip 
We would like to thank Ofer Aharony, Mirjam Cvetic, Koji Hashimoto, 
Hitoshi Murayama, Masaki Shigemori, Herman Verlinde, and
Washington Taylor for discussions. H.O. thanks
the Aspen Center for Physics, where this work was completed. 
 
This research is supported in part by 
DOE grant DE-FG03-92-ER40701. 
Y.O. is also supported in part by the JSPS Fellowship 
for Research Abroad.


\listrefs

\end